# Transparent Compiler and Runtime Specializations for Accelerating Managed Languages on FPGAs

Michail Papadimitriou[a], Juan Fumero[a], Athanasios Stratikopoulos[a], Foivos S. Zakkak[b], and Christos Kotselidis[a]

a   The University of Manchester
b   Red Hat, Inc.

**Abstract**   In recent years, heterogeneous computing has emerged as the vital way to increase computers' performance and energy efficiency by combining diverse hardware devices, such as Graphics Processing Units (GPUs) and Field Programmable Gate Arrays (FPGAs). The rationale behind this trend is that different parts of an application can be offloaded from the main CPU to diverse devices, which can efficiently execute these parts as co-processors. FPGAs are a subset of the most widely used co-processors, typically used for accelerating specific workloads due to their flexible hardware and energy-efficient characteristics. These characteristics have made them prevalent in a broad spectrum of computing systems ranging from low-power embedded systems to high-end data centers and cloud infrastructures.

However, these hardware characteristics come at the cost of programmability. Developers who create their applications using high-level programming languages (e.g., Java, Python, etc.) are required to familiarize with a hardware description language (e.g., VHDL, Verilog) or recently heterogeneous programming models (e.g., OpenCL, HLS) in order to exploit the co-processors' capacity and tune the performance of their applications. Currently, the above-mentioned heterogeneous programming models support exclusively the compilation from compiled languages, such as C and C++. Thus, the transparent integration of heterogeneous co-processors to the software ecosystem of managed programming languages (e.g. Java, Python) is not seamless.

In this paper we rethink the engineering trade-offs that we encountered, in terms of transparency and compilation overheads, while integrating FPGAs into high-level managed programming languages. We present a novel approach that enables runtime code specialization techniques for seamless and high-performance execution of Java programs on FPGAs. The proposed solution is prototyped in the context of the Java programming language and TornadoVM; an open-source programming framework for Java execution on heterogeneous hardware. Finally, we evaluate the proposed solution for FPGA execution against both sequential and multi-threaded Java implementations showcasing up to 224× and 19.8× performance speedups, respectively, and up to 13.82× compared to TornadoVM running on an Intel integrated GPU. We also provide a break-down analysis of the proposed compiler optimizations for FPGA execution, as a means to project their impact on the applications' characteristics.

# The Art, Science, and Engineering of Programming

**Transparent Compiler and Runtime Specializations for Accelerating Managed Languages**

# 1 Introduction

Current systems integrate various compute elements such as CPUs, GPUs, and FPGAs, as a means to offer high performance and energy efficiency. Despite the ongoing efforts to provide high-level programming languages for hardware accelerators, programmability challenges are still present depending on the device. In particular, programming FPGA devices requires a deep understanding of the computing hardware and familiarity with low-level Hardware Description Languages (HDLs) such as Verilog [38] and VHDL [2, 30].

In the last decade, researchers from industry and academia have innovated towards mitigating the steep learning curve of FPGAs' programmability by providing High-Level Synthesis (HLS) [27] tools and heterogeneous programming frameworks (e.g., OpenCL). However, in the realm of managed languages current support for FPGA execution is still very limited. Although a number of JIT compilers that target GPUs have recently emerged [1, 13, 14, 31, 42], in the FPGA domain such solutions cannot be directly applied due to lack of *performance portability* [29, 43] and the necessity to expose low-level hardware primitives to the high-level programming models [36]. To benefit from FPGA acceleration of high-level programming languages, developers must be abstracted away from current FPGA programming norms that require deep hardware understanding and the usage of low-level programming primitives.

In this paper, we describe our experiences from enabling FPGA acceleration of managed programming languages in the context of TornadoVM [12]. We describe the engineering trade-offs we encountered while extending the toolchain with FPGA-acceleration capabilities analyzing the alternative integration paths we implemented for various execution scenarios. In addition, we present our preliminary evaluation results which enacted the implementation of a set of optimizations for specializing the auto-generated FPGA code during compilation.

In detail, this paper makes the following contributions:

- It presents the design and implementation of an open-source end-to-end toolchain designed to transparently compile and run Java code on FPGAs.
- It describes a set of online and offline FPGA execution modes that developers can use depending on the characteristics of their programs for JIT compiling and loading pre-compiled binaries, respectively. Morever, it introduces a complementary emulation mode for fast prototyping.
- It introduces a set of compiler and runtime transformations for specializing the generated FPGA code during compilation.
- It evaluates the TornadoVM FPGA extensions and optimizations on a set of Java benchmarks showcasing end-to-end speedups of up to 19.8×, 224×, and 13.82× over multi-threaded, sequential, and GPU accelerated code, respectively.

■ **Table 1** Taxonomy of the state-of-the-art frameworks that target heterogeneous execution from Java

| Frameworks | Code Generation | Run-time Optim. | HLS Comp. Mode | Hardware Platforms |
|---|---|---|---|---|
| TVM [26] | static | No | No | FPGA Pynq SoC |
| MaxCompiler [25] | dynamic | No | online | Maxeler Platform |
| Aparapi [34] | dynamic | No | offline | AMD GPUs, FPGAs |
| Caldeira et al. [6] | dynamic | Yes | online | Intel Harp FPGAs |
| JOCL [20] | dynamic | No | offline | GPUs |
| TornadoVM [12] | dynamic | No | offline | CPUs, GPUs, FPGAs |
| **TornadoVM + FPGA extensions** | dynamic | Yes | online, offline, emulation | CPUs, GPUs, FPGAs |

## 2 Background

In this section we provide an overview of the current solutions for accelerating managed applications (with a focus on Java) on FPGAs (section 2.1) including the starting point of this work; the original FPGA support of TornadoVM (section 2.2). In addition, it provides an insight on the performance of the initial FPGA integration (section 2.3) which motivated the compiler specializations described in section 4.

### 2.1 Java Execution on FPGAs: Spotting the Gap

Table 1 summarizes the currently available frameworks that enable FPGA acceleration of Java programs. As shown, the frameworks are analyzed based on the following four categories:

**Code Generation** The ability to generate parallel code at compile-time (statically) or at run-time (dynamically). For example, the code generation from Java to OpenCL or Verilog during run-time is classified as dynamic.

**Run-time Optimizations** The ability to automatically specialize code for the target device at run-time, without user intervention (including code annotations).

**HLS Compiler Mode** The ability to perform an online or offline compilation from the generated code to the final FPGA bitstream, via the HLS compilers.

**Hardware Platforms** The supported hardware platforms for FPGA acceleration.

As shown in table 1, the various frameworks offer different functionalities with respect to how they perform the FPGA code generation and code optimizations as well as which platforms they support. In order for FPGA acceleration to become pervasive to the Java programming language, potential solutions must adhere to the core principles of the language including runtime optimizations, dynamic code generation, and hardware-agnostic execution. Based on these principles, we augmented the original TornadoVM with specific FPGA extensions in order to fulfil those requirements as shown in the last line of table 1. The following subsections provide an overview of TornadoVM and its state of FPGA execution prior to the extensions described by this paper.

## 2.2 Background on TornadoVM

TornadoVM enables Java developers to write task-oriented programs that are then automatically compiled and executed on heterogeneous hardware. Each task is essentially a Java method that is automatically compiled to OpenCL at runtime and executed on an OpenCL-compatible device (e.g., multicore CPU or GPU). Originally, the system supported the aforementioned execution flow only for GPUs and CPUs. To enable FPGA execution, developers must manually compile the auto-generated OpenCL code, deploy the generated FPGA bitstream, and redirect the execution from TornadoVM to the FPGA. In addition, the original framework did not perform any compiler optimizations specifically for FPGAs. All the aforementioned limitations prohibited the seamless FPGA execution from within TornadoVM similarly to CPUs and GPUs.

## 2.3 Initial Evaluation

To assess the performance of the original FPGA support of TornadoVM,[1] we performed all the manual steps described in the previous subsection and ran a set of experiments for all benchmarks reported by Fumero, Papadimitriou, Zakkak, Xekalaki, Clarkson, and Kotselidis [12].[2] In almost all cases, we noticed that the achieved performance was slower than the single-threaded Java execution on CPUs. As an example, figure 1 (left) illustrates the relative performance of FPGA execution compared to sequential CPU Java execution[3] when running the Discrete Fourier Transform (DFT) application. As shown, the FPGA execution performs up to 17% slower (for small datasets) than CPUs.

After inspecting the generated code, we noticed that the reason behind this performance degradation is that TornadoVM was tuned and optimized for CPU and GPU acceleration rather than FPGAs. Unlike CPUs and GPUs, FPGAs require hardware-specific annotations to be passed along with the generated OpenCL code in order for the underlying HLS tools to produce an optimal hardware design. To assess the impact of these annotations we revised a proof-of-concept in which we started manually adding OpenCL pragmas to the auto-generated OpenCL kernel. The performance results achieved through this activity are depicted in figure 1 (right). As shown, the manually optimized DFT application outperformed the sequential vanilla Java code executed on the CPU by up to 218 times. These results were in line with the well-documented *performance portability* challenges of OpenCL across different hardware accelerators. Based on our findings, we re-engineered the FPGA execution path of TornadoVM by making its integration seamless with the programming language (section 3) while adding compiler optimization phases to specialize the code for FPGAs automatically (section 4).

---

[1] The exact commit point is: 0093ebcf497f40213dd601c636d906823a050594.
[2] All benchmarks are publicly available at https://github.com/beehive-lab/TornadoVM
[3] See section 5 for the experimental setup.

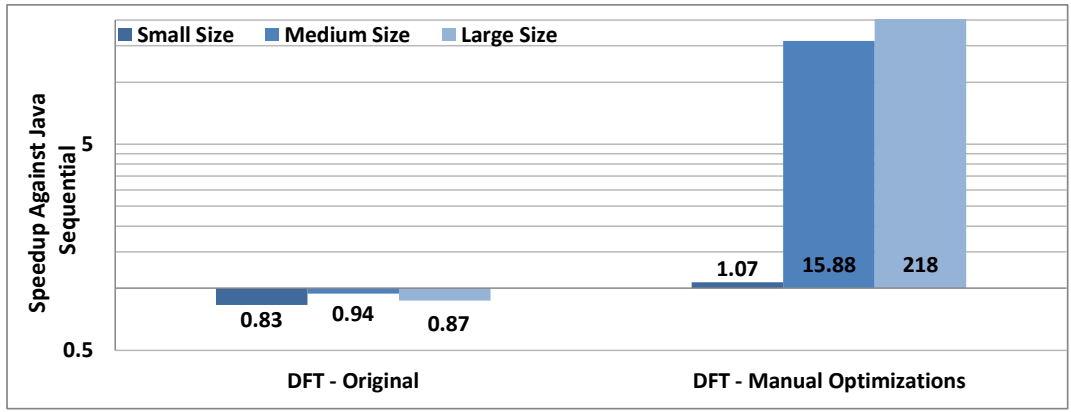

**Figure 1** TornadoVM FPGA results of DFT: a) un-optimized (left), and b) with manual optimizations (right)

## 3 FPGA Extensions in TornadoVM

To address the integration and performance portability challenges, mentioned in section 2, we extended TornadoVM to: a) add support for JIT compilation and emulation mode for seamless execution of Java applications on FPGAs; b) perform a series of automatic compiler optimizations which aim to replace the manual code interventions we performed on the OpenCL generated kernels (section 4); and c) enable users to write a program "once" and "run it anywhere", even on FPGAs, while taking advantage of hardware-acceleration to achieve better performance.

Figure 2 presents the extensions we made to TornadoVM, showcasing how the new approach can be practical for harnessing the FPGA technology within the Java language. The existing components of TornadoVM are illustrated with dark grey, while the applied extensions are depicted in pink.

To generate FPGA code, we extended the TornadoVM's OpenCL backend instead of implementing a new backend for generating HDL similarly to other approaches [3, 23]. Extending the current OpenCL backend to support seamless FPGA execution has the following advantages: a) increasing industrial support and maturity of OpenCL compilers and performance on FPGAs, b) plug-and-play of customized and proprietary bitstream kernels that follow OpenCL semantics in case we do not have access to the source code (legacy or licensed code), and c) it is consistent with the rest of the TornadoVM framework, increasing maintainability.

The remaining of this section describes the individual changes made to the TornadoVM compiler (section 3.1), runtime (section 3.2), and memory management (section 3.3).

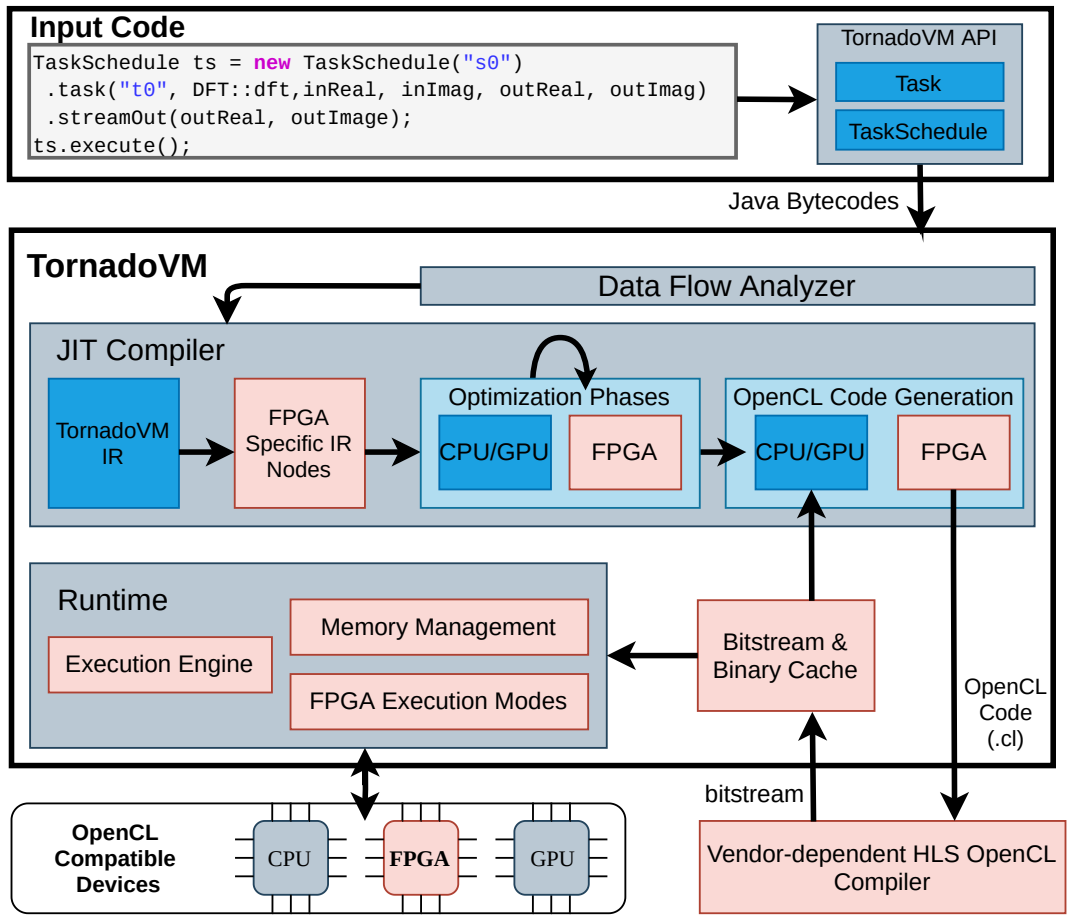

**Figure 2** TornadoVM Overview: the FPGA extensions presented in this work are illustrated in pink.

## 3.1 TornadoVM JIT Compiler

As shown in the work-flow presented in figure 2, the input Java code is compiled to Java bytecodes using the standard Java compiler (javac). Then, the *TornadoVM Data Flow Analyzer* [9] exploits the data dependencies and builds an initial Intermediate Representation (IR) graph of the input program. The generated IR graph is compiled down to the target architecture following the two-stage compilation approach illustrated in figure 3. At the first stage, Java bytecodes are JIT compiled to OpenCL C while at the second stage the OpenCL C code is compiled to FPGA bitstream by the vendors' external toolchains.

During the first-stage compilation, the input IR graph is optimized and specialized through TornadoVM's JIT compiler before the final OpenCL C code emission. Since TornadoVM's JIT compiler is a superset of the Graal [11, 41] compiler, it inherits both its existing set of optimizations and its IR representation. Hence, it employs not only device-specific optimizations and specializations (e.g., for GPUs, multicore CPUs) but also standard compiler optimizations (e.g., loop unrolling, global value numbering, common subexpression elimination, etc.) derived from the Graal compiler.

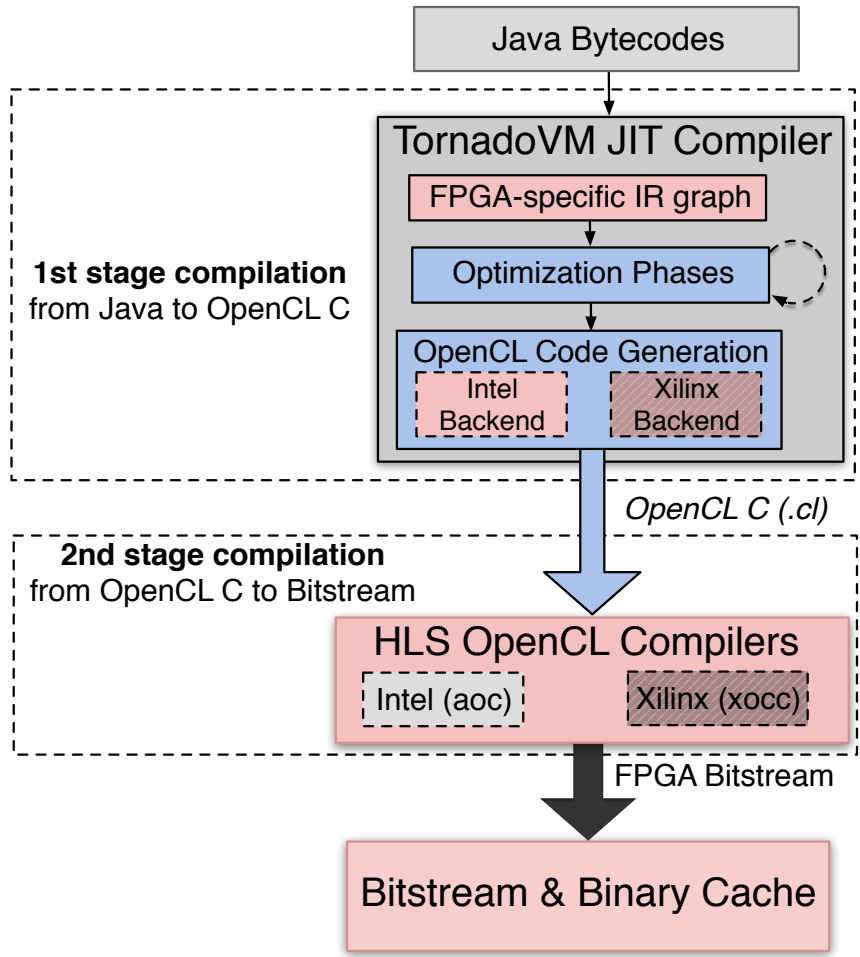

**Figure 3** Two stage compilation: 1) from Java to OpenCL C, and 2) from OpenCL C to FPGA Bitstream

**HLS Integration** After the completion of the first-stage compilation, the generated OpenCL C code is automatically forwarded to the HLS compilers (e.g., Intel's `aoc`); which subsequently perform the second-stage compilation from OpenCL C into the FPGA bitstream. Once the FPGA bitstream is generated, it is stored into the bitstream cache inside TornadoVM. This facilitates the reuse of the bitstreams based on the requirements of the Java programs. Although the current state of the toolchain integrates Intel FPGAs, the proposed system has been designed and implemented in a modular way to support multiple HLS-specific backends. Thus, the toolchain can be extended with insignificant effort for hosting multiple state-of-the-art HLS compilers, such as Vivado HLS from Xilinx.

### 3.2 Extensions to the TornadoVM Runtime

Prior to this work, TornadoVM supported only the ahead-of-time FPGA compilation, which required users to perform the HLS compilation stage manually in the offline mode. In this section we present the various execution modes we added to the runtime that allow programmers to adapt the code execution based on their requirements.

**Execution Modes**  Figure 4 shows the already existed ahead-of-time mode, along with our extensions: the full JIT and emulation modes. The provision of these execution modes allows Java applications to be automatically adapted based on their requirements.

**Ahead-of-Time mode**  This mode alleviates the overhead of the FPGA synthesis process by allowing the plugin of a precompiled bitstream to TornadoVM during execution. The omission of the latency of the second compilation stage (from OpenCL C to bitstream) makes this mode suitable for applications that are sensitive to JIT compilation times (e.g., fast start-up applications or low energy requirements). In addition, since this mode allows users to plug-in their own bitstream implementations, disaggregated machines can be used for FPGA bitstream generation without any limitations or licensing issues.

**Full JIT mode**  This mode enables the end-to-end JIT compilation and execution of Java code onto FPGAs. This is achieved by creating a separate Java thread that makes direct calls to the vendors' HLS compilers for OpenCL (e.g., Intel's OpenCL `aoc` compiler for FPGAs). The HLS compilers for OpenCL follow the traditional process for compiling the OpenCL code into the FPGA bitstream. Once the bitstream is generated, the runtime system stores it into the bitstream cache and marks the Java method *ready* to be executed on the FPGA. Finally, the runtime system loads the generated binary onto the FPGA device, creates the OpenCL program's context, and copies all data required to launch the kernel. In this mode, we enable full JIT compilation from the original Java source code to a fully functional hardware design. However, this JIT compilation process typically requires up to two hours (see section 5) due to the synthesis time required on the FPGA. Thus, the full JIT mode is mainly suitable for long running and server applications, in which the compilation time is offset by the speedups achieved from FPGA acceleration.

**Emulation mode**  The emulation mode is used for fast prototyping, initial debugging, and functional validation of the generated FPGA kernels. This mode is not intended for any performance evaluation, as the emulated kernel code runs on a CPU thread and not on the physical FPGA device. On the contrary, this mode is added to aid developers at the initial stage of development or debugging since it avoids the HLS compilation overheads and it can provide an estimated view of the resource utilization and any compiler warnings associated with the Java code. More importantly, widely available Java IDEs (e.g., Eclipse, IntelliJ, NetBeans) can be used in software development for programming and testing FPGA applications. Hence, developers with no HLS background can experiment by writing pure Java code using standard development tools and assess whether their code can functionally run on an FPGA. The use of standard tools in the development process is also applicable to the other two execution modes.

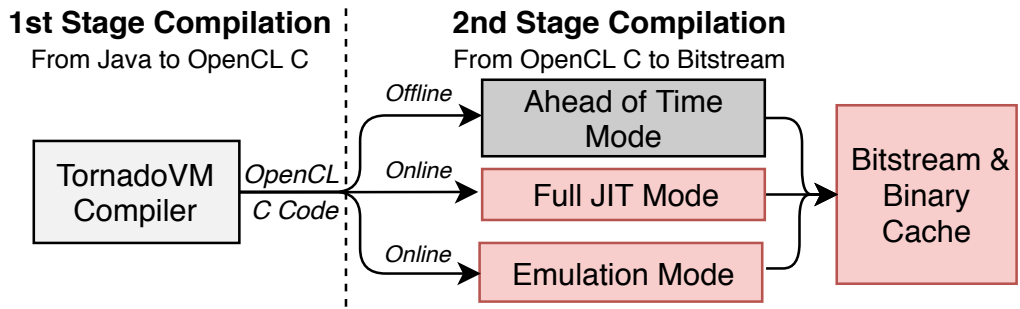

**Figure 4** TornadoVM's execution modes. Our extensions are illustrated in pink.

### 3.3 Memory Management

The memory management between the host and FPGA works as follows: the first time that our framework utilizes an FPGA, it allocates a large amount of device global memory that acts as a managed heap (similarly to a Java heap). The rationale behind this managed on-device heap is to minimize the allocation times on the target device. Our framework performs only a single allocation while performing all data transfers between the host and the FPGA transparently to the user. Furthermore, the proposed toolchain increases the bandwidth between the CPU and the FPGA by using page-locked (or pinned) memory. This enables OpenCL programs to use Direct Memory Accesses (DMA), thereby enhancing the performance of memory transfers. To use pinned memory on the FPGA, we extended the TornadoVM runtime to allocate memory using the OpenCL flag `CL_MEM_ALLOC_HOST_PTR`.[4] The Java stack-frames (memory region that includes the return address and addresses of all input/output buffers on the Java heap) and all Java arrays required for the kernel execution are copied to this allocated region, which is declared as a read/write buffer.

The extended TornadoVM memory manager copies all data to the FPGA's global memory, and it keeps track of all host variables that have been copied to the FPGA. During the runtime data analysis, our toolchain classifies all arrays that are copied to the FPGA as read-only, write-only or read-write. Read-only Java arrays are persisted to the global memory of the device without copying them back to the host's memory. On the contrary, write-only and read-write Java arrays are copied back to the host's memory in order to make their updated values visible to the Java applications. Since TornadoVM dispatches and runs OpenCL code on the FPGA, all operations are, by default, non-blocking. This means that the operations of copy-in, OpenCL kernel launch, and copy-out are non blocking between the FPGA and the main host. Therefore, we added an extra barrier in the TornadoVM bytecode level to wait for the last kernel to be finished before performing the final copy from the device (FPGA) to the host and obtain the results.

---

[4] https://intel.ly/2J0mQFj, last accessed 2020-09-15

■ **Listing 1**    Java snippet for the dft method

```java
private void dft(float[] inreal, float[] inimag, float[] outreal, float[] outimag, int[] inputSize) {
  for (@Parallel int k = 0; k < n; k++) {
    float sumreal = 0;
    float sumimag = 0;
    for (int t = 0; t < n; t++) {
      float angle = ((2 * Math.PI() * t * k) / (float) n);
      sumreal += (inreal[t] * (Math.cos(angle)) + inimag[t] * (Math.sin(angle)));
      sumimag += -(inreal[t] * (Math.sin(angle)) + inimag[t] * (Math.cos(angle)));
    }
    outreal[k] = sumreal;
    outimag[k] = sumimag;
  }
}
```

## 4    Compiler Optimizations for FPGAs

As discussed in section 2.3, although the initial OpenCL-generated code was functionally correct, its performance was not portable on FPGAs. To address this challenge, we introduced a set of compiler optimizations to automatically optimize Java programs for FPGAs without any modification to the user's source code.

### 4.1  Extensions to the JIT Compiler

To enable FPGA-specific optimizations, we extended the Intermediate Representation (IR) of the JIT compiler with FPGA-related nodes. In a nutshell, the compilation flow for FPGAs works as follows: first, the TornadoVM runtime invokes the compiler to build the IR graph that represents the input Java method to be compiled. Consequently, the extended compiler specializes the IR graph for FPGAs through the introduction of new nodes and optimization phases. After the code is optimized and specialized for FPGAs, the final OpenCL C code is generated (figure 4, 1st stage compilation). Finally, the generated code is handled by our extensions to the runtime system, which drives the 2nd stage compilation (figure 4) based on the corresponding execution modes.

The introduced FPGA compiler optimizations are: a) thread-scheduling attributes, b) loop unrolling, and c) loop flattening. Listing 1 shows a Java code snippet for the dft method that we will hereafter use in order to describe the aforementioned optimizations. The listed code contains a method with two nested for loops with the computation residing inside the nested loop. Note that the first loop is annotated (by the developer) using the Java annotation @Parallel, proposed by Clarkson, Fumero, Papadimitriou, Zakkak, Xekalaki, Kotselidis, and Luján [9] to program heterogeneous architectures using TornadoVM. Figure 5 illustrates the compiler transformations that are automatically applied to the IR graph of the code in listing 1. The left-hand side shows the IR graph that corresponds to the code after initially invoking the Tornado compiler. As shown, there are two groups of nodes: data-flow nodes connected by black dashed arrows and control-flow nodes connected by red arrows. Furthermore, each method begins with the Start node. The graph in figure 5-a shows two LoopBegin nodes that correspond to the two loops from the input Java code. To compute the loop bounds of each loop, a phi node along with an if condition is added in the IR. Since our compiler extensions are implemented in TornadoVM, they reuse two new nodes for computing the corresponding indices in OpenCL; namely GlobalID and GlobalSize.

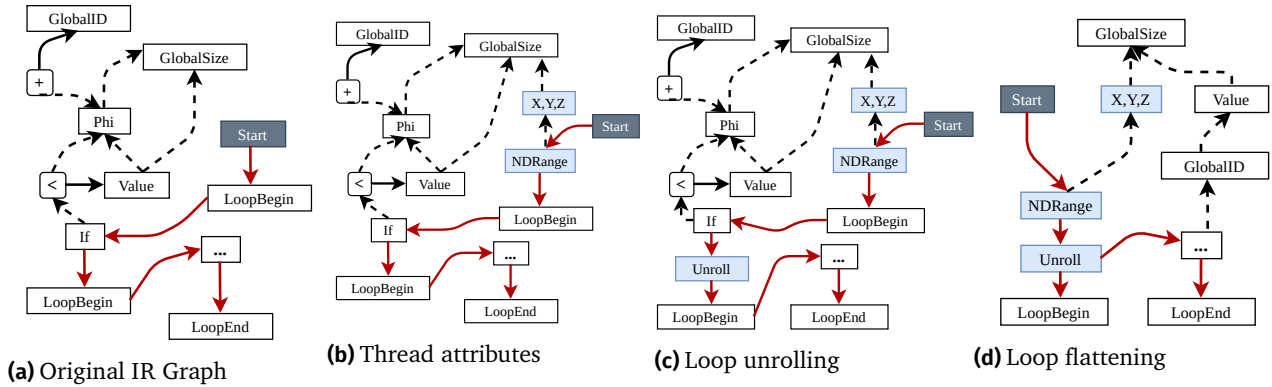

(a) Original IR Graph   (b) Thread attributes   (c) Loop unrolling   (d) Loop flattening

**Figure 5** IR compiler transformations that are automatically performed by our extensions to the TornadoVM's JIT compiler

**Attributes for thread-scheduling** Originally, for kernels targeting OpenCL compatible GPUs, the IR information containing the global indices was sufficient for thread indexing. However, by keeping only the default OpenCL global indices, we ended up generating a kernel designed for single threaded FPGA execution. Consequently, when launching multiple threads on the FPGA, they were pointing to wrong memory locations and thus generating erroneous results. To solve this issue, we extended the IR with a new node for generating OpenCL C attributes before the main kernel. These attributes specify the thread selection (number of threads per block for each dimension — 1D, 2D or 3D) with which the kernel should be executed. This compiler optimization is presented in figure 5-b. A new node called NDRange is inserted right after the Start node. This node points to an additional data-flow node that indicates the values for the thread-blocks in 1D, 2D and 3D (x, y, z) respectively. These values depend on the global size of the compute kernel.

**Loop Unrolling** The second FPGA optimization we apply is loop-unrolling; a widely used optimization for improving OpenCL performance on FPGAs [39]. We analyze whether the inner loop can be unrolled by inspecting the loop bounds. If the input code contains more nested loops, the compiler always tries to perform loop unrolling to the innermost loop. If it can be unrolled, our extended TornadoVM compiler inserts a new control-flow node in the IR before the LoopBegin node of the corresponding candidate loop for unrolling. Figure 5-c highlights this optimization in which the inner loop is annotated by the Unroll node to be unrolled. Our loop unrolling phase uses as a basis the default unroll phase of Graal which means it considers only loops with up to 128 dependency-free iterations as *unrollable*. Consequently, the OpenCL code generator reads the new Unroll node and emits a pragma unroll in the OpenCL C code, leaving the underneath HLS compiler to decide the unrolling factor.

**Transparent Compiler and Runtime Specializations for Accelerating Managed Languages**

**Loop Flattening**  The final compiler optimization that is applied is loop-flattening. The original compiler performs node replacement to substitute for loops with the OpenCL indexing primitives (e.g., get_global_id). However, it maintains the for loops in the OpenCL C code in case the kernel processes more elements than the available threads on the target device. Since our extensions to the TornadoVM compiler specialize the IR for FPGAs, they also specialize the input index space. Therefore, loop nodes can be safely replaced by the OpenCL intrinsic indices. It is important to note that the compiler extensions only flatten the loops that are parallelized, by replacing the loops with the OpenCL indexing primitives. If a loop is computed sequentially, the compiler will preserve the loop nodes. The loop flattening optimization is highlighted in figure 5-d, in which the outermost loop is removed along with every data dependency associated with it. In general, this optimization leads to the simplification of the physical circuits on the FPGA.

### 4.2 Generated FPGA-optimized OpenCL C code

Besides the aforementioned FPGA-targeted optimizations, the extended framework reuses all the compiler optimizations of the original Tornado compiler [9] such as partial escape analysis, dead code elimination, constant propagation, etc. After performing all compiler transformations and optimizations, the toolchain invokes the OpenCL code generator.

Figure 6 provides a sketch of the generated OpenCL code for FPGAs reflecting all described optimizations for the input Java code of listing 1. The left side of figure 6 shows the generated OpenCL code without automatically applying the implemented compiler optimizations, while the right side shows the generated code highlighting the outcomes of the compiler optimizations. The yellow block on the right side highlights the attribute for determining the number of work-items (threads) that is used for thread-scheduling on the FPGA. In our case it is set to 64 elements, as this number has been shown to offer maximum performance on Altera FPGAs [35, 40]. The green-block shows that the outer loop of the orange-block is flattened and the remaining loop is only indexed by the *get_global_id* OpenCL intrinsic. This optimization simplifies the generated hardware circuits on the FPGA and, therefore, increases performance. Finally, the blue-block highlights the loop unrolling for the FPGA with a factor of two through the *pragma unroll* OpenCL Intel FPGA directive before the innermost loop.

Once the OpenCL FPGA code is generated, we call the underlying OpenCL compiler (e.g., the Intel `aoc` compiler) and compile the OpenCL C source code to the FPGA bitstream as explained in section 3.

### 4.3 Current Limitations

Although the extended framework provides a seamless way to compile, execute and rapidly prototype Java code on FPGAs, it still poses some limitations that prevent it from achieving higher performance. Examples include the support for fine-grain memory, such as private and local, and the exploitation of advanced OpenCL features such as pipes for direct intra-kernel communication.

```
__kernel void compute(__global uchar *_heap_base,
                      ulong _frame_base, … ) {
// variable declaration
...
__global ulong *_frame=(__global ulong*)
&_heap_base[frame_base];

base0 = (ulong) _frame[6];
base1 = (ulong) _frame[7];
base2 = (ulong) _frame[7];
tid   = get_global_id(0);
...
i8 = *((__global int *) &_heap_base[base0]);
for(;tid < maxElements) {
  ...
  f10 = 0.0F;
  i11 = 0;
  for(;i11 < i8;) {
    ...
  }
  ul_38 = base1 + index;
  *((__global float *) &_heap_base[ul_38]) =
  ul_37 = base2 + index;
  *((__global float *) &_heap_base[ul_39]) =
  i_40 = get_global_size(0);
  i_41 = i_40 + tid;
  tid = i_41;
}
```

```
// Scheduling attributes
__attribute__((reqd_work_group_size(64,1,1)))
__kernel void compute(__global uchar *_heap_base,
                      ulong _frame_base, … ) {
// variable declaration
...
__global ulong *_frame = (__global ulong *)
&_heap_base[_frame_base];

base0 = (ulong) _frame[6];
base1 = (ulong) _frame[7];
base2 = (ulong) _frame[7];
tid   = get_global_id(0);              // Loop flattening
...
i8 = *((__global int *) &_heap_base[base0]);
...
f10 = 0.0F;
i11 = 0;
#pragma unroll 2      // Loop unrolling with factor 2
for(;i11 < i8;) {
  ...
}
ul_38 = base1 + index;
*((__global float *) &_heap_base[ul_38]) = result1;
ul_37 = base2 + index;
*((__global float *) &_heap_base[ul_39]) = result2;
}
```

**Figure 6** Sketch of the generated OpenCL code specialized for FPGAs

## 5 Evaluation

### 5.1 Experimental Methodology

We evaluate the performance of the FPGA executed code of our toolchain against the *peak* performance of single and multi-threaded Java implementations compiled with the server compiler (C2) of OpenJDK [28]. In addition, we performed a comparative evaluation of the FPGA accelerated code against an Intel HD Graphics 630 integrated GPU. In order to guarantee that the JVM has been warmed up, we perform up to 50 iterations per benchmark and then we report the mean of the consequent 10 runs. To ensure the functional correctness of the generated FPGA code (section 3.2), we validated all benchmarks in all execution modes.

The time for the FPGA-executed code is also reported using the mean of 10 runs, similarly to the CPU-executed code. Furthermore, all reported numbers correspond to end-to-end executions, which include the times for loading the bitstreams into the FPGA, executing the kernels, copying the data from the main host to the FPGA memory, and copying back the data from the FPGA memory to the host (CPU). Finally, we evaluate each benchmark against three different workloads—small, medium, and large—with data sizes increasing by an order of magnitude, varying from 1 MB to 540 MB. The size of the large workloads corresponds to the maximum size permitted by the HLS compiler for mapping each generated circuit on the FPGA device.

■ **Table 2** Array length and the input/output data sizes of the benchmarks

| Benchmark | Array Length | | | Input (MB) | Output (MB) |
|---|---|---|---|---|---|
| | Small | Medium | Large | Large | Large |
| VectorAdd | 32768 | 1048576 | 67108864 | 540 | 268 |
| Grayscale | 32768 | 1048576 | 33554432 | 268 | 140 |
| BlackScholes | 256 | 1048576 | 33554432 | 268 | 536 |
| RenderTrack | 64 | 1024 | 8192 | 268 | 200 |
| N-Body | 256 | 16384 | 32768 | 6 | 3 |
| DFT | 64 | 65536 | 262144 | 2 | 1 |

#### 5.1.1 Benchmarks

For the evaluation[5] we use two standard benchmarks (VectorAdd and BlackScholes), two variations of computational dwarfs (NBody and DFT), and two computationally intensive kernels for image processing (RenderTrack and Grayscale). For all benchmarks we provide both sequential and multi-threaded Java implementations that have been ported and verified using various data sizes. Table 2 presents the actual lengths of the arrays that correspond to the three workloads (small, medium and large) used for evaluating each benchmark. Moreover, we present the size of the input/output data for the large workloads in *Megabyte (MB)*.

#### 5.1.2 Experimental Setup

The results presented in this section are conducted on a computer system consisting of an Intel i7-7700K CPU, clocked at 4.20GHz, featuring 64GB of RAM and a Nallatech 385A Accelerator Card attached via PCI-e. The accelerator card features an Intel Arria 10 FPGA (*10AX115N3F40E2SG*) and two banks of DDR3 SDRAM with 4GB each. The system runs CentOS 7.4 with Linux kernel 3.10. In addition, the Arria 10 FPGA offers native IEEE 754 single-precision floating-point operations through its DSP blocks [17]. We use OpenCL 1.0 with Intel FPGA SDK 17.1 and the OpenCL HPC Board Support Package (BSP). Note that the FPGA frequency for all the kernels is automatically determined by Intel's OpenCL compiler and ranges from 176 to 218 MHz. For all experiments we used the Java OpenJDK 1.8.0_131 64-Bits (C2 compiler) with the Java Virtual Machine Compiler Interface (JVMCI) [6] enabled and 16GB of Java heap memory.

### 5.2 Performance Analysis

We evaluate the proposed system in terms of execution speedup over three different Java execution scenarios. The first two concern the performance acceleration over sequential and multi-threaded Java execution, while the last one compares the specialized FPGA execution against an Intel integrated graphics card.

---

[5] All benchmarks are publicly available at https://github.com/beehive-lab/TornadoVM.
[6] https://openjdk.java.net/jeps/243

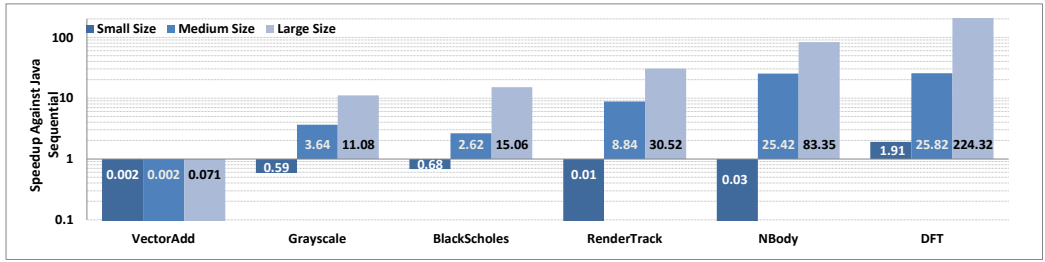

**Figure 7** Speedup of Intel Arria 10 FPGA against sequential Java for small, medium and large data sizes

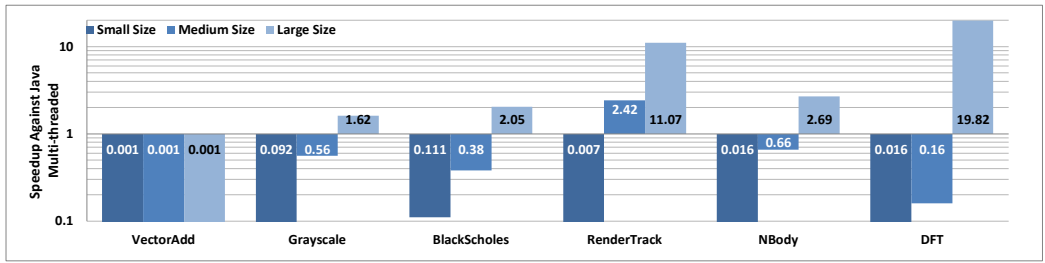

**Figure 8** Speedup of Intel Arria 10 FPGA against multithreaded Java (8 threads) for small, medium and large data sizes

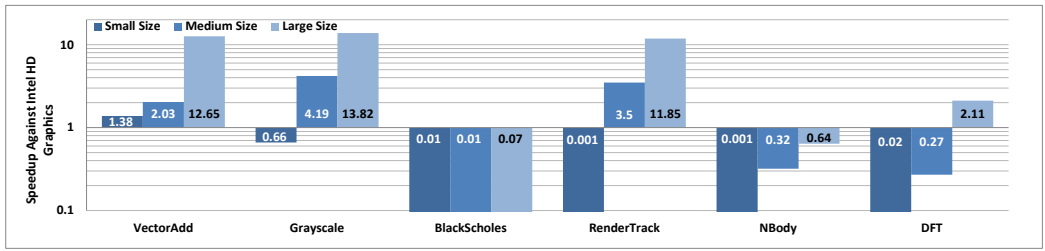

**Figure 9** Speedup of Intel Arria 10 FPGA against Intel HD Graphics 630 for small, medium and large data sizes

**FPGA versus sequential Java code**    Figure 7 shows the performance of the FPGA executed code against the sequential Java code. As shown, for small workloads FPGA execution exhibits performance slowdowns across all benchmarks except for DFT. This is due to the fact that the time spent in data transfers is significantly higher than the time of the FPGA computation. In the case of memory-bound benchmarks such as VectorAdd, the performance slowdown can reach up to $0.002x$. Regarding Grayscale, BlackScholes and RenderTrack, although they perform lower compared to sequential Java for small workloads, they show performance scalability while increasing data sizes, with peak speedups of $11x$, $15x$ and $30x$ respectively. Moreover, NBody shows similar behaviour with a peak speedup of $83x$. Finally, for highly computational benchmarks, such as DFT, the FPGA outperforms the CPU-executed sequential Java code for all input sizes by up to $224x$.

**Transparent Compiler and Runtime Specializations for Accelerating Managed Languages**

**FPGA versus multithreaded Java code**    Figure 8 shows the performance of the FPGA executed code against the multithreaded Java code. All benchmarks utilize the maximum number of available threads in the system (eight), except RenderTrack for which *Hyper-Threading* was deactivated since it was resulting in performance degradation. As shown in figure 8, for large data sizes FPGA execution outperforms the multithreaded Java implementations from $1.62x$ up to $19.82x$. However, for small and medium input data sizes, the multi-threaded Java code outperforms the FPGA executed code with the exception of the RenderTrack benchmark. Again, this is due to the overhead of copying data to and from the FPGA.

**FPGA versus an Intel HD Graphics 630 GPU**    Figure 9 shows the performance of the FPGA executed code against TornadoVM running on an Intel HD Graphics 630 integrated GPU. The local- workgroup configuration is not manually tuned, instead the local_work_size attribute is left empty for the clEnqueueNDRangeKernel. Therefore, the Intel driver, and its OpenCL implementation will automatically determine how to split the global work-items. As shown in figure 9, for all the benchmarks except NBody and Blackscholes, for large data sizes, FPGA execution outperforms the Intel Graphics card up to $13.82x$. For medium-sized workloads performance varies depending on the application. However, for small sizes the FPGA always performs worse than the Intel HD Graphics card due to the overheads which occurs for copying the data to/from the device. This behaviour is expected and for such reason it is expected to use the FPGA for acceleration in cases where highly dense workloads are present.

### 5.2.1 Runtime Overhead Analysis

To further understand the performance of the system, we performed a breakdown analysis of the end-to-end execution times of all benchmarks, as presented in figure 10. We analyze the execution times only for the largest input data sizes in order to highlight the impact of the data transfers between the host and the device memories. Each bar has four parts which correspond to: a) kernel execution time on the FPGA (*Kernel*), b) data transfer time from host to device (*H2D*), c) data transfer time from device to host (*D2H*), and d) the (*Rest*). The *Rest* includes the time for loading the binary on the FPGA and initializing the OpenCL context of each kernel.

Figure 10 shows that up to 18% is spent in transferring data from the host to the FPGA device (*H2D*) and backwards (*D2H*). In particular, VectorAdd, BlackScholes and Grayscale spent up to 10 %, and RenderTrack up to 18 % of their total time in data transfers. On the contrary, the *Kernel* execution time is up to 99 % for both computationally intensive benchmarks; NBody and DFT.

The VectorAdd benchmark is a special case because it exhibits slowdowns, as illustrated in figure 7, even though the kernel execution percentage is large enough to anticipate performance improvements. The reason is that this benchmark is memory intensive and the current version of the toolchain does not support memory-specific optimizations (section 4.3) that could increase its performance. Examples of such optimizations are auto-vectorization support for load and store operations in combination with local memory. Finally, the time for loading the binary and initializing OpenCL contexts (*Rest*) across all benchmarks is negligible.

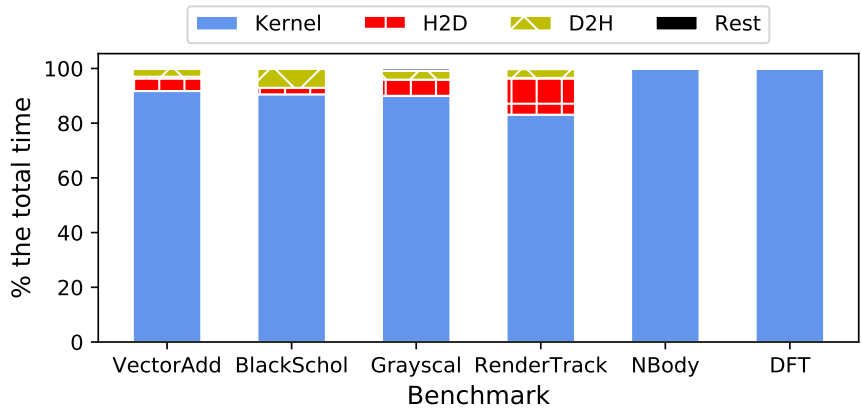

**Figure 10** Execution time breakdown

In a nutshell, our findings in figures 7 to 9 show that due to the selection of compute-intensive benchmarks, the performance scales while increasing the input size. The only exception is for `VectorAdd`, in which the Java single-threaded (figure 7) and multi-threaded (figure 8) implementations outperform the FPGA execution for all data sizes. However, as shown in figure 9, the performance of the FPGA execution also scales for different data sizes, when comparing against the Intel integrated GPU. Since the majority of the time for all benchmarks is spent during the kernel execution, the cost to copy data from the host to the device, and backwards, is not a performance bottleneck.

#### 5.2.2 Optimization Phases Breakdown

Table 3 presents a breakdown analysis for the contribution of each optimization phase to the overall performance. We show the performance obtained after applying the optimizations in three phases. The first phase includes only the thread scheduling (TS) optimization and shows two different thread-block configurations; one with 32 threads and another one with 64 threads.[7] As shown, performance increases by up to 35 times and 56 times when applying thread-scheduling with block sizes 32 and 64, respectively.

The second phase applies loop unrolling (LU) optimization on top of TS. By performing this combination (TS_64 + LU), performance increases up to 214 times (3.8 times improvement). Finally, if we combine the previous optimizations with loop flattening (TS 64 + LU + LF), performance increases by up to 224 times.

### 5.3 HLS Compilation & Binary Loading

Table 4 shows the HLS compilation times, binary loading times, and the sizes of the generated FPGA bitstreams for each benchmark. The HLS compilation time regards the 2nd stage compilation (section 3.2) and it is the time spent for compiling the

---

[7] The selection of the thread-block configurations was taken after conducting a series of test-runs with varied sizes and selecting the best performing ones.

**Transparent Compiler and Runtime Specializations for Accelerating Managed Languages**

■ **Table 3** The impact of each optimization phase in performance. The first optimization phase includes Thread-Scheduling (TS), the second phase applies Loop Unrolling (LU) along with scheduling with 64 threads (TS_64). The final phase includes all previous optimizations and Loop Flattening (LF).

| Benchmark | TS | | TS_64 + LU | TS_64 + LU + LF |
|---|---|---|---|---|
| | TS_32 | TS_64 | | |
| VectorAdd | 0.0002 × | 0.0001 × | NA | 0.07 × |
| Grayscale | 9.42 × | 10.01 × | NA | 11.08 × |
| BlackScholes | 15.85 × | 15.97 × | NA | 15.06 × |
| RenderTrack | 21.23 × | 28.16 × | NA | 30.52 × |
| NBody | 11.04 × | 31.20 × | 66.75 × | 83.35 × |
| DFT | 35.44 × | 56.31 × | 214.64 × | 224.32 × |

■ **Table 4** Bitstream size, loading, and HLS times

| Benchmark | Bitstream Size (MB) | Load Bitstream (ms) | HLS Compilation (minutes) |
|---|---|---|---|
| VectorAdd | 172 | 22 | 48 |
| Grayscale | 173 | 23 | 52 |
| BlackScholes | 174 | 24 | 54 |
| RenderTrack | 173 | 23 | 44 |
| NBody | 173 | 24 | 114 |
| DFT | 173 | 22 | 68 |

specialized generated OpenCL C code to bitstream. The binary loading time is the time required to load the bitstreams and initialize the OpenCL context on the FPGA on behalf of the running program. As shown, while the binary loading time is in the range of milliseconds, the HLS compilation time can take up to 114 minutes to complete. Furthermore, the HLS compilation time includes the timing for placement and routing, which is a process strongly related to the vendor tools and the complexity of the generated kernels. For instance, the NBody kernel reports the longest compilation time as it includes the loop unrolling optimization which utilizes more private memory on the FPGA and thus higher BRAM resources (table 5). The increased latency in HLS compilation times was the motivation for providing a set of execution modes in TornadoVM that can either perform a whole compilation for FPGAs at runtime (Full JIT), or load the bitstream of pre-compiled kernels (AOT). Nevertheless, the OpenCL drivers for Xilinx and Intel are evolving quickly, thereby reducing the compilation time and making JIT compilation more affordable [18]. Finally, as shown in table 4, both binary loading times and bitstream sizes are consistent among the benchmarks.

■ **Table 5** Resource utilization as reported by AOC

| Benchmark | LUTs | | FFs | | DSPs | | BRAM | |
|---|---|---|---|---|---|---|---|---|
| VectorAdd | 145 535 | (19.5 %) | 275 521 | (18.4 %) | 72 | (3 %) | 570 | (39.4 %) |
| Grayscale | 117 928 | (15.7 %) | 230 040 | (15.4 %) | 72 | (3 %) | 494 | (34.0 %) |
| BlackScholes | 186 348 | (24.9 %) | 306 361 | (20.5 %) | 490 | (20.7 %) | 935 | (64.7 %) |
| RenderTrack | 118 582 | (15.9 %) | 238 742 | (16 %) | 72 | (3 %) | 514 | (35.5 %) |
| NBody | 174 036 | (23.3 %) | 329 764 | (22.1 %) | 120 | (5.1 %) | 1291 | (89.3 %) |
| DFT | 146 418 | (19.6 %) | 264 652 | (17.7 %) | 109 | (4.6 %) | 748 | (51.7 %) |
| Resources | 747 080 | | 1 494 160 | | 2367 | | 1446 | |

**5.4 Resource Utilization**

Table 5 shows the FPGA's resource utilization of four different hardware components—Look Up Tables (LUTs), Flip Flops (FFs), Digital Signal Processing (DSPs), and Memory Blocks (BRAM)—for each benchmark. As shown, the utilization of the LUTs varies between 15.9 % and 24.9 % of the total capacity of the FPGA. In particular, BlackScholes utilizes more LUTs and DSPs than the rest of the benchmarks, as the generated OpenCL C code contains 216 lines of code with complex control flow. The utilization of DSPs is between 3 % and 5.1 % for all benchmarks, with the exception of BlackScholes which is at 20.7 % again due to its code complexity. Regarding BRAM utilization, NBody is a special case occupying up to 89.3 % of the available resources because our extensions to the TornadoVM JIT compiler unrolls two of the innermost loops.

Overall, the results indicate that the current set of benchmarks utilizes roughly one fourth of the available resources on the FPGA, except BRAMs. BRAMs show higher utilization due to loop unrolling, which duplicates the memory accesses and the intermediate stored values.

**5.5 Discussion on the Suitability of FPGA Acceleration for Java Applications**

By carefully analyzing the benchmarks and the obtained results of FPGA execution, we revised a set of technical guidelines to answer the question of when FPGA acceleration is suitable for Java applications.

**Applications not suitable for FPGAs** Applications such as VectorAdd in which the runtime needs to copy a significant amount of data to compute just a few operations (one in the case of VectorAdd) are not suitable for FPGA execution. This is due to the fact that modern CPUs which operate at much higher frequency than FPGAs (GHz versus MHz), can perform a larger number of such operations in less time. In more detail, the FPGA operates at up to 300 MHz while the CPU up to 4.2 GHz. In addition, the Java HotSpot compiler makes use of high-performance vector instructions and operations (e.g., fused multiply-add (FMA)).

**Applications suitable for FPGAs** Applications such as BlackScholes, Grayscale, NBody, DFT and RenderTrack exhibit significant speedups when operating over large data sizes.

**Transparent Compiler and Runtime Specializations for Accelerating Managed Languages**

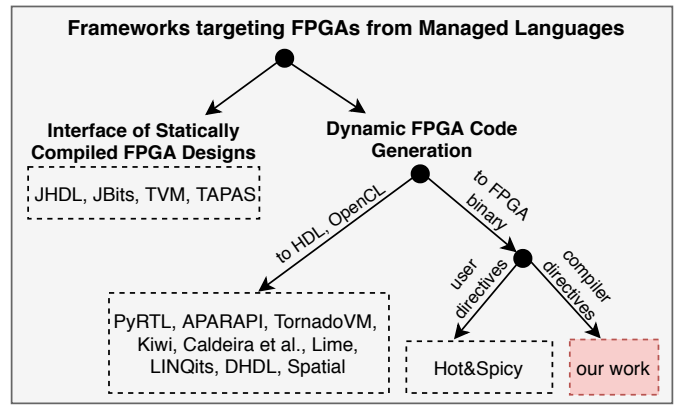

**Figure 11** Classification of the state-of-the-art frameworks that target FPGA hardware from managed languages

These applications make use of the specialized hardware on the FPGA for accelerating the abundance of math operations they contain (e.g., sine and cosine). This is due to the fact that FPGAs can perform these operations in few clock cycles. Although, this class of applications exhibits speedups for large datasets, it is expected that their performance will increase if further memory optimizations are implemented in TornadoVM. Nevertheless, the trade-off between computational acceleration and data transferring overheads will be present for all applications.

## 6 Related Work

The prior work on FPGA acceleration of managed languages (e.g., Java, Python, C#) can be classified into the two categories illustrated in figure 11. The first category regards languages that interface statically with pre-compiled FPGA designs, while the second category includes languages that generate FPGA code dynamically.

### 6.1 Interfacing with static FPGA designs

Bellows and Hutchings [5] introduced the JHDL framework which describes how a Java interface should be constructed for accessing the FPGA hardware by calling existing bitstreams. Guccione, Levi, and Sundararajan [16] presented the JBits API that requires hardware knowledge in order to leverage hardware acceleration, assuming that a compiler can be used to automate code generation without providing the necessary information.

Additionally, several frameworks have been proposed for applications written in domain specific languages (DSL) that provide bindings or build on top of other managed languages. These frameworks compile the DSL applications into hardware units that can be accelerated at runtime on top of general purpose FPGAs. Moreau, Chen, and Ceze extended TVM [7, 26], an optimizing compiler for deep learning applications, to utilize FPGAs. Margerm, Sharifian, Guha, Shriraman, and Pokam [24] presented the TAPAS framework that analyzes task dependencies in the compiler graph and generates a data-flow processing unit that is transparently executed on the FPGA.

**6.2 Dynamic FPGA code generation**

Several frameworks that enable FPGA-based acceleration for managed languages have been recently introduced. Clow, Tzimpragos, Dangwal, Guo, McMahan, and Sherwood [10] presented the PyRTL framework which compiles Python programs onto Verilog. However, it requires programmers to be familiar with specific design practices and hardware primitives. Fumero, Papadimitriou, Zakkak, Xekalaki, Clarkson, and Kotselidis [12] and Segal, Margala, Chalamalasetti, and Wright [34] extended the GPU capabilities of two Java-based frameworks TornadoVM and APARAPI, respectively, to run on OpenCL compatible hardware. However, the proposed frameworks did not support any automatic optimization phases and several kernels required manual intervention in order to be synthesized on FPGA boards. Greaves and Singh [15] presented the Kiwi library that exposes various custom attributes to the programmers, and generates Verilog HDL from the C# input code. Caldeira, Penha, Braganca, Ferreira, Nacif, Ferreira, and Pereira [6] presented a framework that compiles Java programs into Verilog HDL. However, further interpolation is required to imprint the resulting Verilog code into the FPGA of Intel HARP platforms. In addition, Skalicky, Monson, Schmidt, and French [36] proposed Hot&Spicy to compile code written in a subset of Python into HLS C code and transparently invoke the Xilinx SDSoC HLS tool to produce the FPGA binary.

Furthermore, several approaches have proposed other domain specific languages (DSLs) (e.g., Lime [3], Language Integrated Query (LINQ) [8], Delite Hardware Definition Language (DHDL) [23], Spatial [22]) to generate code for various FPGA hardware description languages [21]. Auerbach, Bacon, Cheng, and Rabbah [3] presented the Lime framework within the streaming domain that compiles programs to Java, C and Verilog. Chung, Davis, and Lee [8] proposed the LINQits framework that allows various Big Data workloads to be compiled by the Dandelion [33] compiler and accelerated on FPGA hardware. However, LINQits does not support automatic HLS compilation and requires programmers to introduce the HLS directives. Koeplinger, Prabhakar, Zhang, Delimitrou, Kozyrakis, and Olukotun [23] also required users to write their program into the DHDL language which subsequently compiled into MaxJ; a low-level Java-based language that allows the generation of hardware for the Maxeler platform with the usage of the MaxCompiler. In addition, Koeplinger, Feldman, Prabhakar, Zhang, Hadjis, Fiszel, Zhao, Nardi, Pedram, Kozyrakis, and Olukotun [22] proposed the Spatial language and compiler, as an extension to DHDL, thereby allowing developers to gain more control over the memory hierarchy from the programming language.

Moreover, a couple of frameworks leverage the abstraction that compiler intermediate representations (IRs) provide to dynamically generate HLS-compatible input. Sozzo, Baghdadi, Amarasinghe, and Santambrogio [37] introduced FROST, a unified backend for targeting FPGAs for DSLs, such as Halide [32] and Tiramisu [4]. FROST provide its own IR along an IR optimizer with FPGA-oriented passes and a scheduling co-language to allow users to specify optimizations. Izraelevitz, Koenig, Li, Lin, Wang, Magyar, Kim, Schmidt, Markley, Lawson, and Bachrach [19] presented FIRRTL a Flexible Intermediate Represe ntation for RTL. FIRRTL is integrated with Chisel which is a hardware design language that facilitates advanced circuit generation and design reuse for both ASIC and FPGA digital logic designs. In addition, it transforms target-independent RTL into design-specific RTL through a number of optimization steps such as simplifying transformations, analyses, optimizations, instrumentations, and specializations.

To the best of our knowledge Hot&Spicy [36] is the only framework that accepts a program written in a managed language (i.e., Python) and produces an FPGA binary. Nonetheless, Hot&Spicy requires the programmer to add hardware-specific primitives in Python; and transforms the input program to contain the appropriate wrapper bindings for interfacing with the generated hardware design. Our work differs from all aforementioned frameworks since it: a) automatically and dynamically compiles Java programs onto optimized FPGA binary code, b) it does not require the use of hardware-specific directives from programmers, and c) it is able to accelerate existing Java applications without any modifications.

# 7 Conclusions

This paper presents a practical approach that augments managed languages with the ability for seamless and efficient FPGA code execution. It presents the engineering challenges and trade-offs when integrating the different toolchains for achieving end-to-end JIT compilation of Java code to FPGA bitstreams. In addition, we showcase how specific specialization and optimization phases can be transparently added in order to increase the performance of unoptimized FPGA code. We prototyped our approach in the context of TornadoVM, by introducing a two-stage compilation process and a set of FPGA-specific specialization techniques. We evaluated the proposed framework against a set of Java benchmarks executed on an Intel FPGA showcasing speedups up to 19.8, 224, and 3.82 times over multi-threaded, sequential, and GPU-accelerated Java code, respectively.

In the future we plan to further extend our work to enable automatic use of private and local memory of FPGAs, and to enhance the compiler in order to exploit more advanced OpenCL features, such as channels and pipes.

**Acknowledgments**

This work is partially supported by the EU Horizon 2020 E2Data 780245 grant.

**Transparent Compiler and Runtime Specializations for Accelerating Managed Languages**

## About the authors

**Michail Papadimitriou** is a PhD Candidate at the University of Manchester. His research is focused on JIT compilation, heterogeneous architectures and parallel programming. Contact Michail at michail.papadimitriou@manchester.ac.uk.

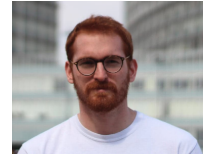

**Juan Fumero** is a Research Associate at the University of Manchester. His research is focused on heterogeneous computing, parallel programming models, parallel architectures and hardware accelerator for managed programming languages. Contact Juan at juan.fumero@manchester.ac.uk.

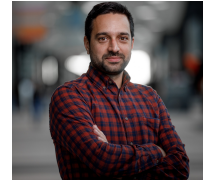

**Athanasios Stratikopoulos** is a Research Associate at the University of Manchester working on heterogeneous architectures and reconfigurable accelerators. His interests include computer architecture, high-performance computing, storage systems and hardware acceleration in the cloud. Contact Athanasios at athanasios.stratikopoulos@manchester.ac.uk.

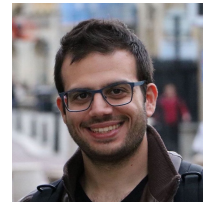

**Foivos S. Zakkak** is a Software Engineer at Red Hat, Inc. The work presented in this article was performed while employed by The University of Manchester. His work focuses on managed runtime systems, parallel programming, and performance optimizations. Contact Foivos at fzakkak@redhat.com.

**Christos Kotselidis** is a Senior Lecturer of Computer Science at the University of Manchester, UK. He is currently leading the Beehive Lab which specializes in system software in the areas of hardware acceleration of managed languages and hardware/software co-design. Contact Christos at christos.kotselidis@manchester.ac.uk.

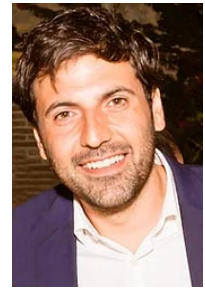